\documentclass[reprint, superscriptaddress, secnumarabic, amssymb, nobibnotes, aps, prl]{revtex4-1}
\usepackage{epstopdf} 
\usepackage{amsbsy}
\usepackage{gensymb}
\usepackage{adjustbox}
\usepackage{siunitx}
\usepackage[T1]{fontenc}
\usepackage{amsmath}
\usepackage{amssymb}
\usepackage{bbm}
\usepackage{xcolor}
\usepackage{graphicx}
\usepackage[colorlinks=true]{hyperref}
\hypersetup{
    bookmarks=true,
    unicode=false,
    pdftoolbar=true,
    pdfmenubar=true,
    pdffitwindow=false,
    pdfstartview={FitH},
    pdftitle={x},
    pdfauthor={SM},
    pdfsubject={},
    pdfcreator={},
    pdfproducer={},
    pdfkeywords={},
    pdfnewwindow=true,
    colorlinks=true,
    linkcolor=blue,
    citecolor=blue,
    filecolor=magenta,
    urlcolor=blue
}
\usepackage[normalem]{ulem}

\renewcommand{\approx}{\simeq}

\newcommand{\sample}{Nb$_{0.25}$Ta$_{0.25}$Ti$_{0.25}$Zr$_{0.25}$}

\begin{document}

\preprint{AIP/123-QED}

\title{Multiband Superconductivity and High Critical Current Density in Entropy Stabilized \sample}

\author{Nikita Sharma}
\affiliation{Department of Physics and Material Science, Thapar Institute of Engineering and Technology, Patiala 147004, India}
\author{Kuldeep Kargeti}
\affiliation{Department of Physics, Bennett University, Greater Noida-201310, Uttar Pradesh, India}
\author{Neha Sharma}
\affiliation{Department of Physics and Material Science, Thapar Institute of Engineering and Technology, Patiala 147004, India}
\author{Pooja Chourasia}
\affiliation{Department of Physics and Material Science, Thapar Institute of Engineering and Technology, Patiala 147004, India}
\author{B. Vignolle}
\affiliation{Univ. Bordeaux, CNRS, Bordeaux INP, ICMCB, UMR 5026, Pessac, F-33600, France}
\author{Olivier Toulemonde}
\affiliation{Univ. Bordeaux, CNRS, Bordeaux INP, ICMCB, UMR 5026, Pessac, F-33600, France}
\author{Tirthankar Chakraborty}
\affiliation{Department of Physics and Material Science, Thapar Institute of Engineering and Technology, Patiala 147004, India}
\author{S. K. Panda}
\email[]{swarup.panda@bennett.edu.in}
\affiliation{Department of Physics, Bennett University, Greater Noida-201310, Uttar Pradesh, India}
\author{Sourav Marik}
\email[]{soumarik@thapar.edu}
\affiliation{Department of Physics and Material Science, Thapar Institute of Engineering and Technology, Patiala 147004, India}
\affiliation{Univ. Bordeaux, CNRS, Bordeaux INP, ICMCB, UMR 5026, Pessac, F-33600, France}

\date{\today}

\begin{abstract}
High and medium-entropy superconductors with significant intrinsic disorder are a fascinating class of superconductors. Their combination of robust structural integrity, superior mechanical properties, and exceptional irradiation tolerance makes them promising candidates for use in advanced superconducting technologies. Herein, we present a comprehensive theoretical and experimental investigation on the superconductivity of equiatomic entropy-stabilized \sample. The material shows bulk superconductivity (transition temperature = 8K) with a high upper critical field of 11.94T. Interestingly, both the electronic band structure and specific heat data point toward unconventional multiband superconductivity. Our ab initio calculations reveal Dirac-like band crossings close to the Fermi level, with certain degeneracies persisting even in the presence of spin-orbit coupling, suggesting a possible interplay between topological electronic states and the observed unconventional superconductivity. Remarkably, the critical current density exceeds the benchmark of 10$^5$A/cm$^2$, surpassing all previously reported as-cast entropy-stabilized superconductors. This high critical current density is likely attributed to strong flux pinning at the grain boundaries, facilitated by extreme intrinsic lattice distortion. Taken together, the demonstrated dynamical stability, excellent metallicity, potential to host unconventional superconductivity, and exceptionally high critical current density highlight the potential of entropy-stabilized alloys as a platform for exploring the confluence of disorder, topology, and unconventional superconductivity.

\end{abstract}

\maketitle

\section{Introduction}
Quantum materials, an exotic class of material systems, have attracted significant research attention due to their promising applications in quantum computing, advanced energy storage, and quantum sensing technologies. In this connection, exploration of novel unconventional superconductors has emerged as a very crucial research topic in recent times. Their unique ability to sustain quantum coherence and support qubit formation, the fundamental building blocks of quantum information, places them at the core of future quantum devices. Besides, superconductors are also integral to a range of important applications, for instance, nuclear magnetic resonance (NMR) spectroscopy, magnetic resonance imaging (MRI), fusion reactors, and high-energy particle accelerators \cite{1, 2}.

Recently, the emergence of superconductivity in high entropy stabilized materials has opened new directions in superconducting research \cite{3, 4, 5}. The first observation of superconductivity in a multicomponent high entropy alloy (HEA) Ta$_{34}$Nb$_{33}$Hf$_{8}$Zr$_{14}$Ti$_{11}$ (superconducting transition temperature $T_C = 7.3$ K, upper critical field, $H_{c2}(0) = 7.75$ T), was reported in 2014 and has since drawn attention for its possibility to realize unconventional superconductivity, including the topological one \cite{5}.  Remarkably, despite this huge inherent disorder, HEAs have demonstrated robust superconductivity \cite{5, 6, 7}, alongside exceptional mechanical and radiation-tolerant properties, such as exceptional mechanical performance at high and low temperatures \cite{8, 9, 10}, oxidation resistance \cite{11}, and resistance to irradiation \cite{12}. All these make them highly attractive for demanding applications in aerospace, radiation-rich environments (fusion reactor), and advanced superconducting systems. 

Since the discovery of superconductivity in HEA, subsequent investigations into both high entropy and medium-entropy materials have revealed a variety of structural configurations exhibiting intriguing superconducting properties. For instance, hexagonal close-packed (HCP) structure is realized in Re$_{0.56}$Nb$_{0.11}$Ti$_{0.11}$Zr$_{0.11}$Hf$_{0.11}$ with T$_{C}$ = 4.4 K \cite{13}, CsCl structure is observed in superconducting (ScZrNb)$_{1-x}$(RhPd)$_x$ \cite{14}. High upper critical field is observed in complex A15-type Nb$_3$(AlSnGeGaSi) \cite{15}. Possible unconventional superconductivity with topological feature is observed in Ti$_{0.2}$Nb$_{0.2}$Ta$_{0.2}$Zr$_{0.2}$Hf$_{0.2}$C \cite{4}. NiAs type (RuRhPdIr)$_{1-x}$Pt$_x$Sb demonstrates the possibility to improve the superconducting properties utilizing high chemical disorder \cite{16}. Work on the high-entropy alloy (HEA) thin films has highlighted a remarkable increase in the critical current density (J$_C$) in comparison to the as-prepared bulk form, making them highly suitable for both superconducting devices and large-scale applications \cite{12}. Notably, these films display exceptional resilience to irradiation-induced disorder.  

The field of entropy-stabilized disordered superconductors is still in its infancy, offering exciting opportunities to explore unconventional forms of superconductivity. However, most of the previous studies have highlighted relatively low superconducting transition temperatures, typically below 7 K. In this work, we present a comprehensive theoretical and experimental investigation of a NbTi-based medium-entropy alloy, \sample, which exhibits a superconducting transition temperature of 8 K, the highest among medium and high-entropy alloys to date. The presence of tantalum (Ta) in the structure is expected to introduce strong spin-orbit coupling (SOC), potentially serving as a key ingredient for realizing unconventional superconductivity. We have shown the possibility of multiband superconductivity, high upper critical field, and high J$_C$ exceeding 10$^5$ A/cm$^2$ in the material. Our theoretical calculations highlight the possibility of unconventional superconductivity in the material. 

\section{Methodology}
\subsection{Synthesis method}
The equiatomic as-cast high-quality polycrystalline medium entropy alloy \sample \ was synthesized using the vacuum arc melting technique. The constituent elements, niobium pieces (purity 99.9 $\%$), tantalum slug (purity 99.95 $\%$),  titanium slug (purity 99.98 $\%$) and zirconium pieces (purity 99.9 $\%$)  were precisely weighed and melted in a Ti-gettered ultra-high-purity Ar atmosphere. The alloy was melted at high temperature and then flipped and melted several times to achieve chemical homogeneity. A small piece of as-cast sample is annealed at 550 $\degree C$ for 6 hours with Argon flow. 
\subsection{Structural and microstructural characterizations}
The structural characterization was done using a Rigaku diffractometer with Cu-K${\alpha 1}$ radiation ($\lambda = 1.5406$ \text{\AA}). The microstructural characterization was analyzed using a ZEISS GEMINI field emission scanning electron microscope (FE-SEM) and energy dispersive x-ray spectroscopy (EDS) mappings were obtained using a BRUKER XFlash 6160 system.
\subsection{Magnetic and physical property measurements}
The magnetic measurement was performed utilizing a superconducting Quantum Interference Device (SQUID) magnetic property measurement system (MPMS-XL7 and MPMS-3). The temperature-dependent magnetic behavior of the material was investigated in both field-cooled (FC) and zero-field-cooled (ZFC) modes. To investigate further, the magnetic hysteresis (M-H) loops were taken at various temperatures to elucidate the magnetic response of the material. The transport and thermodynamic measurements were studied using the Quantum Design Physical Property Measurement System (QD-PPMS). The electrical resistivity measurements were performed as a function of temperature across a range of applied magnetic fields ranging from 0 to 10 T. The temperature dependence of specific heat was also measured under various magnetic fields ranging from 0 to 9 T.  
\subsection{Computational details}
To investigate the atomic-scale electronic structure of the compositionally disordered high-entropy alloy (HEA) \sample, we employed the Special Quasirandom Structure (SQS) methodology to effectively model the random solid solution inherent to HEAs. The SQS approach provides a computationally tractable representation of disordered alloys by generating a finite supercell that closely mimics the statistical correlation functions of an ideal random alloy. To construct the SQS, we utilized a Monte Carlo-based simulated annealing algorithm as implemented in the \texttt{mcsqs} module of the Alloy Theoretic Automated Toolkit (ATAT)~\citep{17, 18, 19}. This algorithm optimizes both the supercell geometry and atomic site occupancies to reproduce pair and higher-order correlation functions, ensuring a statistically representative model of the disordered HEA. 

Electronic structure calculations were conducted within the framework of density functional theory (DFT) using the Vienna Ab initio Simulation Package (VASP)~\citep{20, 21, 22}. The exchange-correlation interactions were described using the Perdew-Burke-Ernzerhof (PBE) functional within the generalized gradient approximation (GGA)~\citep{23}. To ensure high numerical precision, a plane-wave basis set with an energy cutoff of 550 eV was employed.  To obtain a reliable equilibrium configuration, full structural relaxation was performed, allowing simultaneous optimization of atomic positions, cell volume, and lattice parameters. The relaxation process continued until the residual forces on each atom were reduced to below $10^{-3}$eV/\AA, ensuring a highly accurate equilibrium geometry suitable for subsequent analyses.

Phonon spectra were computed using density functional perturbation theory (DFPT) as implemented in VASP. This approach enables the calculation of vibrational properties by evaluating the dynamical matrix through linear response theory, providing insights into the lattice dynamics and stability of the \sample. The DFPT calculations were performed on the fully relaxed structure, ensuring consistency with the optimized geometry.

\section{Results and Discussion}
\begin{figure*}
\centering
\adjustbox{width=2.0\columnwidth, center}
{\includegraphics{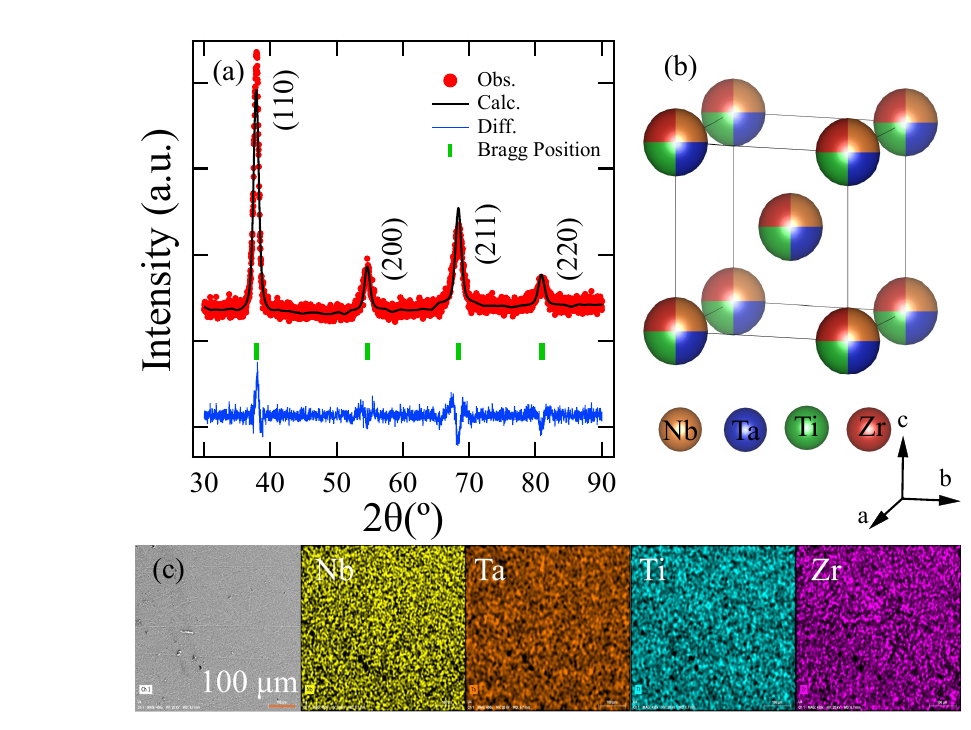}}
\caption{\label{Fig1:xrd}(a) The room temperature powder X-ray diffraction pattern (red sphere) and Rietveld refinement (black line) of \sample .\ (b) Schematic representation of BCC lattice with randomly distributed atoms. (c) The FESEM micrograph with EDS elemental mapping of Nb, Ta, Ti, Zr.}
\end{figure*}
The room temperature powder XRD pattern for the \sample \ is presented in Figure \ref{Fig1:xrd}(a). We analyzed the recorded XRD pattern using the FULLPROF suite software. The Rietveld refinement of the material confirms a body-centered cubic (BCC) crystal structure with the space group Im-3m, showing no impurity peaks. The peaks are labelled with their respective Miller indices and the refined lattice parameters are a = b = c = 3.364 (1) \text{\AA}. The atomic radii for Nb, Ta, Ti, and Zr are 1.429 \text{\AA}, 1.43 \text{\AA}, 1.46 \text{\AA}, and 1.60 \text{\AA}, respectively. Utilizing the rule of mixtures (Vegard's law), the theoretical lattice parameter of the solid solution is calculated to be 3.365 \text{\AA} \cite{24}. It is important to note that the variation in atomic radii among the constituent elements has resulted in the broadening of the XRD peaks for the material. Figure \ref{Fig1:xrd}(b) depicts the crystal structure of the cubic BCC sample, where the atoms arrange themselves on the crystallographic positions that exhibit high configurational entropy. The FESEM micrographs of the \sample \, material are shown in Figure \ref{Fig1:xrd}(c) along with the elemental mappings. Clearly, the elemental distribution of the studied material is even throughout. The quantitative chemical analysis performed at various positions revealed that the atomic compositions for each element are approximately 25 $\%$ close to the design value (provided in table S1 in the supplemental information (SI) file \cite{25}). To further confirm the phase stability of the material, various parameters based on the composition are evaluated and summarized in Table 1.

\begin{table}[h!]
\caption{Parameters for phase identification of \sample.}
\label{Parameters}
\begin{center}
\begin{tabular*}{1.0\columnwidth}{l@{\extracolsep{\fill}}lll}\hline\hline
Parameters& Criteria (single phase)& \sample \\ \hline                               
$\Delta S_{mix} (J/K)$& $>0.69$R& $1.38$R\\ 
$\Delta H_{mix} (kJ/mol)$& $\approx 0$& $0.5$\\ 
$\Omega$& 1.1$< \Omega <$229.8& $58.12$\\ 
$\Delta\chi$& & $0.10$\\ 
$\delta (\%)$& < 6.6& $4.82$\\
$VEC$& < 6.87 (BCC)& $4.5$\\ 
\hline\hline
\end{tabular*}
\par\medskip\footnotesize
\end{center}
\end{table}

\begin{figure*}[t]
\centering
\adjustbox{width=2.0\columnwidth, center}
{\includegraphics{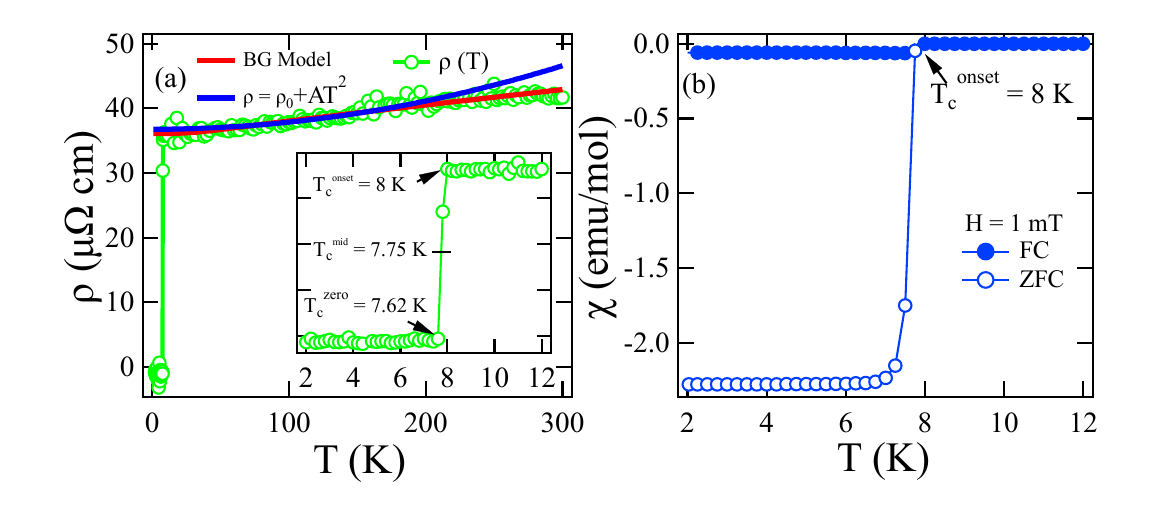} }
\caption{\label{Fig2:MT}(a) The temperature-dependent dc-electrical resistivity (H = 1 mT) along with the Bloch-Gruneisen (BG) and power law fitting in the range of 2 K - 300 K for \sample.\ The inset shows the expanded low-temperature superconducting transition state. (b) Temperature variation of the dc-magnetic susceptibility (H = 10 Oe) measured in FC and ZFC mode. Both the measurements highlight the onset of the superconducting transition at 8 K.}
\end{figure*}
In order to understand the electrical transport properties and superconducting and normal state properties of the \sample, the resistivity, $\rho \, (T)$ measurements were taken from 2 K to 300 K in the absence of a magnetic field, as shown in Figure \ref{Fig2:MT}(a). The inset in Figure \ref{Fig2:MT}(a) shows that the resistivity sharply drops to zero at $T_C^{\text{onset}}$ = 8 K and reaches a zero-resistive state at $T_C^{zero}$ = 7.62 K. The mid-point of the normal state resistivity and the zero resistive state is considered as the transition temperature, $T_C^{mid}$ = T$_C$ = 7.75 K. The residual resistivity ratio, RRR ($\rho_{300}/\rho_{10}$), is a measure of the quality of the material and is estimated as 1.16. This low value of RRR indicates the existence of high-atomic-scale disorder and significant defect scattering at low temperature. This value of RRR is comparable to the reported values of other HEAs and MEAs \cite{26, 27}.
The low-temperature resistivity behavior was well fitted using a power law dependence of resistivity,
\begin{equation}
    \rho = \rho_{0} + A T^n
\end{equation}
which yield residual residual resistivity $\rho _0 = 36.3 \,  \mu \, \ohm \, cm$ and low-temperature electron scattering coefficient, A = $1.09 \times 10^{-4}\,\mu\,\ohm \, cm \,K^{-2}$ and n $=$ 2 consistent with the Fermi-liquid theory. To analyse further, the normal state resistivity of the \sample \,we considered the Bloch-Gruneisen (BG) model \cite{28}. According to this model, the resistivity is given by, 
\begin{equation}
    \rho (T) = \rho _0 + \rho _{ph}(T)
\end{equation}
where, $\rho_{ph}(T)$ is phonon-mediated resistivity and is given by the expression:
\begin{equation}
\rho_{ph}(T) =  C\left(\frac{T}{\theta_D}\right)^n \int_0^{\theta_D/T} \frac{x^n}{(e^x-1)(1-e^{-x})}\, dx    
\end{equation}
In this equation, C is a material-dependent constant related to electron-phonon coupling, and n depends on the scattering process, $\theta_D$ is the Debye temperature. The value of n = 3 provide best fit to the data and allows extraction of $\theta_D = 206.9$ K. The Kadowaki-woods ratio parameter, $A/\gamma_n^2 = 0.47 \times 10^{-5}\,\mu\, \ohm \, cm \,K^2\,mJ^{-2}\,mol^2$ classifies the electron-electron interaction as weakly correlated. Here, the parameter $\gamma_n$ is the normal state Sommerfeld coefficient determined from the specific heat measurement as shown in Figure \ref{Fig5:Spheat}(a)  

\begin{figure*}
\centering
\adjustbox{width=2.0\columnwidth, center}{\includegraphics{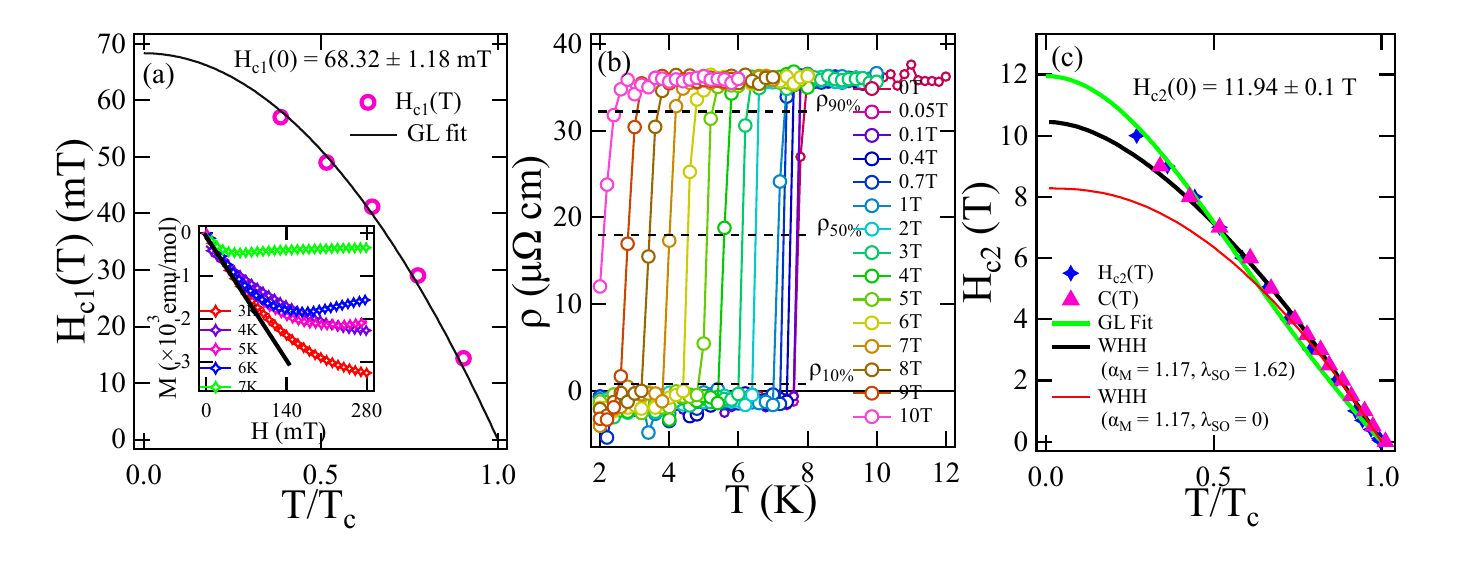}}
\caption{\label{Fig3:Resistivity} (a) The observed lower critical field as a function of reduced temperature. Inset shows the field-dependent magnetization for the temperature range 3 K to 7 K (b) The temperature-dependent resistivity transition to zero resistance up to 10 T. (c) Upper critical field calculated from resistivity and specific heat measurements analyzed with GL model and WHH model \cite{29, 30}.}
\end{figure*}

The bulk superconducting transition for the \sample \, investigated from the ZFC and FC dc magnetization measurement (Figure \ref{Fig2:MT}(b)), shows a similar superconducting onset temperature to the resistivity measurement. The magnetization as a function of applied field measurements performed at different temperatures was used to estimate the value of the lower critical field ($H_{c1}(0)$). The magnetization varies linearly with the applied magnetic field, but after a certain value, it starts deviating from linearity as the field increases. This deviation of magnetization is taken as the $H_{c1}$ for different temperatures. The obtained values of $H_{c1}(T)$ with normalized temperature are shown in Figure \ref{Fig3:Resistivity}(a). The lower critical field at absolute zero, $H_{c1}(0)$, is extrapolated as 68.32 mT using Ginzburg-Landau (GL) approximation:
\begin{equation}
  H_{c1}(T) = H_{c1}(0)\left[1-\left(\frac{T}{T_C}\right)^2\right]  
  \label{eqn1:hc1}
\end{equation}

The upper critical field, $H_{c2}(0)$, is an important parameter for estimating the material performance and in exploring the potential practical applications of the superconductors in magnetic fields. In order to determine $H_{c2}(0)$, electrical resistivity measurements were taken in varying magnetic fields. It is evident from the Figure \ref{Fig3:Resistivity}(b) that the increasing value of the magnetic field suppresses T$_C$. The mid point of the resistivity drop, $T_C^{mid}$, is considered as the criterion of transition temperature for extrapolating the value of $H_{c2}(0)$. The $H_{c2}(T)$ as a function of temperature is plotted in Figure \ref{Fig3:Resistivity}(c). The experimental data are modelled by the GL expression,
\begin{equation}
    H_{c2}(T) = H_{c2}(0) \frac{1 - (T/T_\mathrm{c})^2}{1+(T/T_\mathrm{c})^2}
\end{equation}
The GL model provides a reasonable fit to the experimental data for the entire temperature range. The extrapolated value of $H_{c2}(0)$ is 11.94 T. Furthermore, $H_{c2}(0)$ is directly related to the Ginzburg-Landau coherence length ($\xi_{GL}(0)$), by the relation \cite{31}:
\begin{equation}
\label{eqn1:coherence}
    H_{c2}(0) = \frac{\phi_0}{2\pi\xi_\mathrm{GL}^2(0)}
\end{equation}
where $\phi_0$ is the magnetic flux quantum number ($h/2e$) having the value $2.07 \times 10^{-15}$ \text{T $m^2$}. By using the value of $H_{c2}$(0), we have obtained $\xi_{GL}$(0) = 52.4 \AA. The magnetic penetration depth $\lambda_{GL}(0)$ is associated with the $H_{c1}(0)$ and $\xi_{GL}(0)$ according to the expression \cite{32}
\begin{equation}
    H_{c1}(0) = \frac{\Phi_{0}}{4\pi\lambda_\mathrm{GL}^2(0)}\left(\ln\frac{\lambda_\mathrm{GL}(0)}{\xi_\mathrm{GL}(0)}+0.12\right)
\end{equation}
We have obtained $\lambda_{GL}(0) = 2850$ \AA \, by substituting the values of $H_{c1}(0)$ and $\xi_{GL}(0)$ obtained from the equation \ref{eqn1:hc1} and equation \ref{eqn1:coherence}. The ratio of magnetic penetration depth and coherence length ($\lambda_{GL}(0)/\xi_{GL}(0)$) is $54.389 >> 1/\sqrt{2}$, favoring the \sample \, as a type-II superconductor.

In type-II superconductors, the superconductivity is completely suppressed after $H_{c2}$(0).  The breaking of Cooper pairs into electrons is due to the action of the Lorentz force acting on the paired electrons, which have opposite momenta. The limiting case arises when the kinetic energy of the paired electrons exceeds the condensation energy of the Cooper pairs. For a BCS superconductor, the orbital limiting upper critical field $H_{c2}^{orbital}$(0) is given by the Wertham$-$Helfand-Hohenberg (WHH) expression \cite{29, 30}:
\begin{equation}
    H_{c2}^{\text{orbital}}(0) = -\alpha T_C \left. \frac{dH_{c2}(T)}{dT}\right|_{T=T_C}
\end{equation}
 where $\alpha = 0.693$ is the purity factor defined for dirty-limit superconductors. The slope $-\frac{dH_{c2}}{dT}$ at T = T$_C$ is 1.87. With this value, the WHH model yielded the $H_{c2}^{orbital}(0) = 10.04$ T. The Pauli paramagnetic field of a type-II superconductor is given by, $H_{c2}^{Pauli} = 1.85 \, T_C = 14.33$ T. The values obtained from the GL model and the WHH model are smaller than the $H_{c2}^{Pauli}$. In order to analyze the influence of orbital limiting field and Pauli paramagnetic effect, the Maki parameter ($\alpha_M$) is calculated \cite{33, 34}. The obtained value of $\alpha_M$ is $\approx$ 1 using the relation
 \begin{equation}
     \alpha_M = \sqrt{2}\frac{H_{c2}^{orbital}(0)}{H_{c2}^{pauli}}
 \end{equation}
 
\begin{figure*}
  \centering
\adjustbox{width=2.0\columnwidth, center}
    {\includegraphics{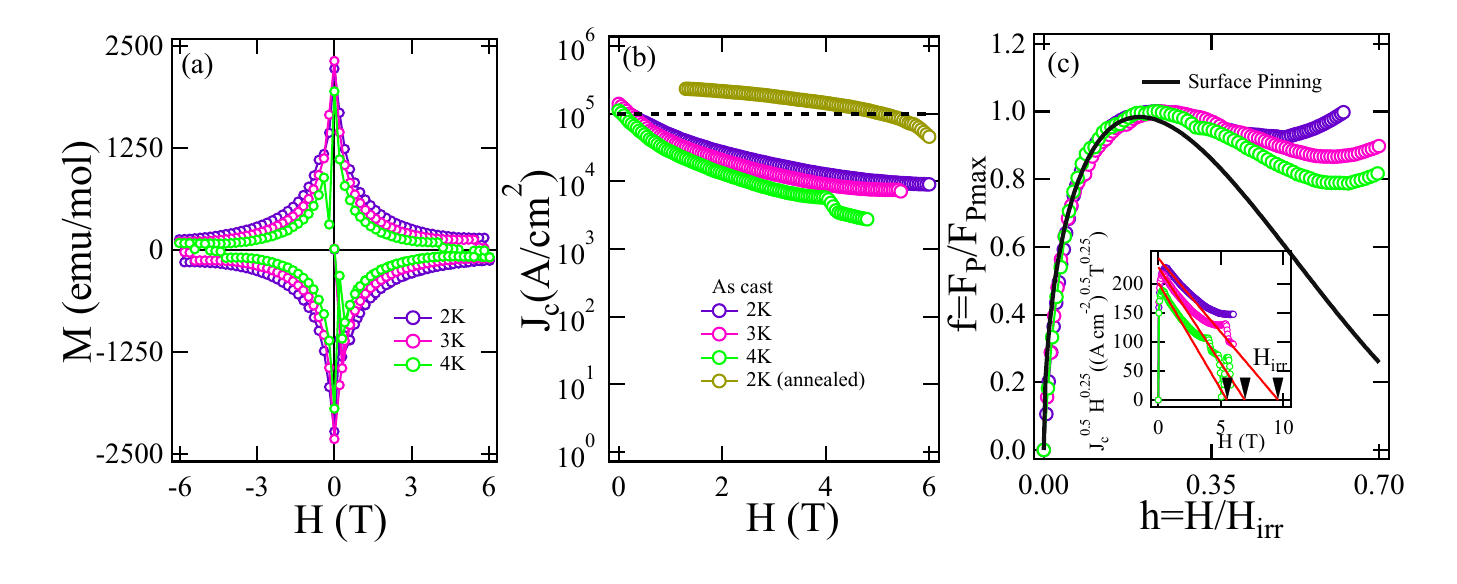}}
    \caption{(a)Magnetic hysterisis loop (M-H) at temperatures ranging from 2 K to 4 K for \sample. (b) Magnetic field dependence of critical current density ($J_C$) plotted in logarithmic scale. A very high $J_C$ value, exceeding the practical benchmark of $10^{5} \text{A/cm}^2$ is observed. The $J_C$ exceeds all the previously reported as-cast high and medium-entropy alloys by a few orders of magnitude. (c) Normalized flux pinning density as a function of reduced field. The inset shows the Kramer plot to estimate the irreversible field. }
    \label{fig4:MH}
\end{figure*}

\begin{figure*}
\centering
\adjustbox{width=2.0\columnwidth, center}
{\includegraphics{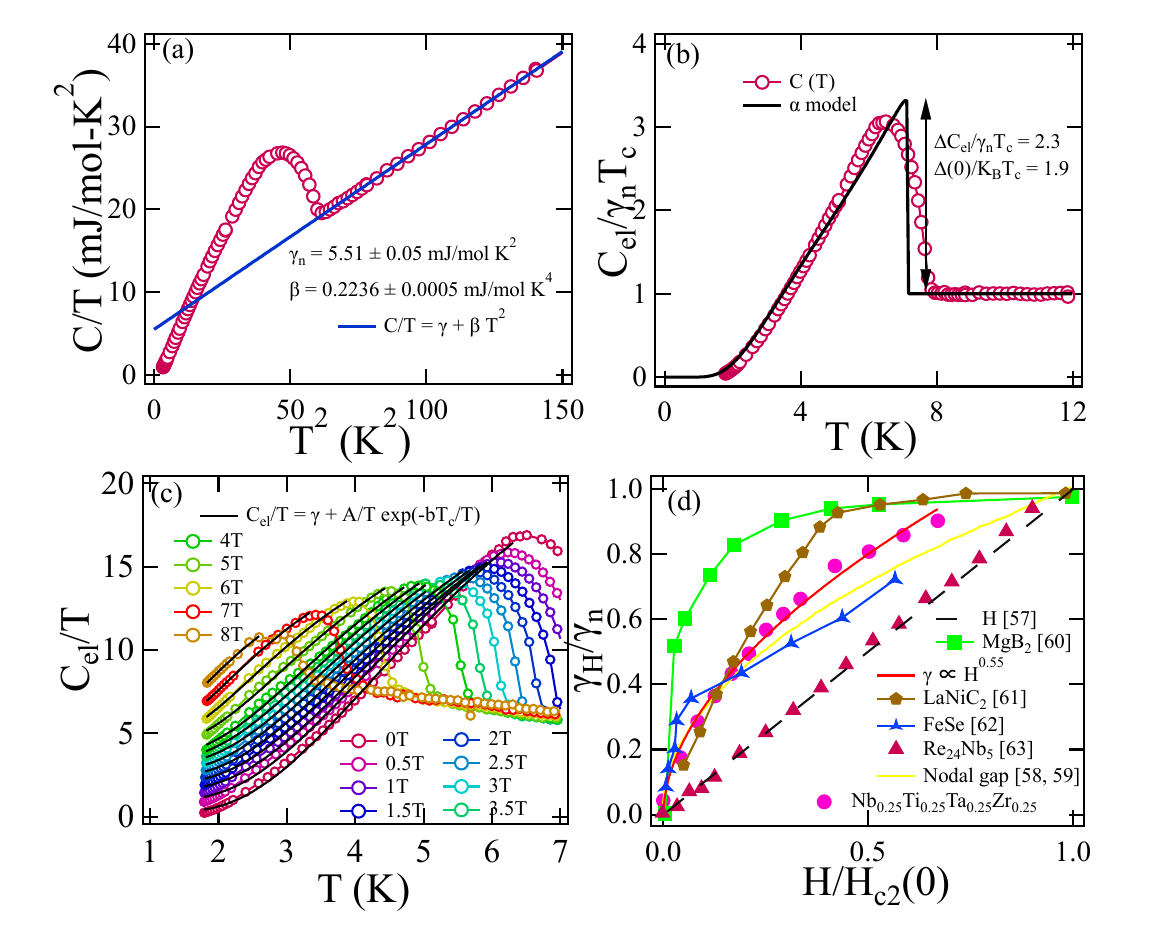}}
\caption{\label{Fig5:Spheat} (a) C/T vs T$^2$ in zero field for \sample, fitted using Debye model to extract electronic and phononic contribution. (b) Normalized electronic specific heat as a function of temperature. The specific heat data is fitted with the $\alpha$ model and highlights a very high specific heat jump. (c) Field dependent electronic specific heat with temperature. The data is fitted using $\frac{C_{el}}{T} = \gamma + \frac{A}{T}exp (-\frac{bT_C}{T})$. (d) Normalized Sommerfeld coefficient ($\gamma _H/ \gamma _n$) as a function of reduced field (H/H$_{c2}$(0)). The behaviour of the $\gamma _H/ \gamma _n$ highlight the possibility of multiband superconductivity.}
\end{figure*}
 
 This near equality signifies a balanced relationship between the two pair-breaking mechanisms. This explains the discrepancy between the $H_{c2}(0)$ values obtained from the GL fit and the WHH fit. This near equality underscores the need to include spin-orbit coupling ($\lambda_{SO}$) in our analysis based on one-band WHH model that includes spin paramagnetism and the spin-orbit scattering parameter, in addition to the orbital limiting effect \cite{29, 30, 35}. 
 \begin{align}
\begin{aligned}
\ln \left(\frac{1}{t}\right) = \left(\frac{1}{2} + \frac{i \lambda_{SO}}{4 \gamma}\right) \psi\left(\frac{1}{2} + \frac{\bar{h} + \lambda_{SO}}{2} + \frac{i \gamma}{2t}\right)\\
+ \left(\frac{1}{2} - \frac{i \lambda_{SO}}{4 \gamma}\right) \psi\left(\frac{1}{2} + \frac{\bar{h} + \lambda_{SO}}{2} - \frac{i \gamma}{2t}\right) - \psi\left(\frac{1}{2}\right)
\label{eqn3:WHH}
\end{aligned}
\end{align}
where:
\[
\gamma \equiv \left(\alpha \bar{h}\right)^2 - \left(\frac{\lambda_{SO}}{2}\right)^2,
\]
and
\[
h^* \equiv \left. \frac{d\bar{h}}{dt} \right|_{t=1} = \frac{\pi^2 \bar{h}}{4} = \left. \frac{dH_{c2}}{dt} \right|_{t=1}
\]

where, $t = \frac{T}{T_C}$ is the reduced temperature, $\bar{h}$ is the reduced field, $\psi$ is the digamma function and $h^*$  represents the reduced field's \( \bar{h} \) initial slope with respect to the reduced temperature \( t \) at \( t = 1 \) (i.e., near \( T_C \)). 
At first, we assumed the absence of the spin-orbit scattering effect (i.e., $\lambda_{SO} = 0$). The curve deviates significantly from the experimental data. The experimental fitted data provides us ($\lambda_{SO}$ = 1.63), which appropriately describes the dependence of $H_{c2}$ of the studied \sample. A significant spin-orbit scattering factor is responsible for enhancing the upper critical field of the sample.

One of the most critical parameters for the practical application is the critical current density ($J_C$). A $J_C$ of 10$^5$ A/cm$^2$ is often considered a practical benchmark for technologies such as high-field superconducting magnets \cite{1}. We have utilized Bean's model to calculate the $J_C$. From Beans model, $J_C$ is expressed as follows \cite{36}:
\begin{equation}
    J_C = \frac{2 \Delta M}{a(1-a/3b)}
\end{equation}
In this equation, $\Delta M = M_{up}-M_{down}$ represents the vertical width of the magnetic hysteresis loop measured in the units of A/cm (in terms of magnetization per unit volume ($a\times b \times c$), where c $<$ a $<$ b are the dimensions of the sample. Figure \ref{fig4:MH}(a) shows the magnetic hysteresis loop at temperatures ranging from 2 K to 6 K for the as-cast \sample. The magnetic field dependence of the measured $J_C$  is presented in Figure \ref{fig4:MH}(b). The measured value of $J_C$ is $\approx$ 0.13 $\times\,10^{6} \text{A/cm}^2$ for all three temperatures below 0.25 T magnetic field. This value exceeds the $J_C$ values of all the previously reported as-cast high- and medium-entropy alloys by a few orders of magnitude. For instance, the as-cast arc melted MEA sample Nb$_{0.4}$Hf$_{0.2}$Zr$_{0.2}$Ti$_{0.2}$ shows a $J_C$ value of 0.6$\times \,10^{4} \text{A/cm}^2$ \cite{37}. As-cast arc melted HEA sample Ta$_{1/6}$Nb$_{2/6}$Hf$_{1/6}$Zr$_{1/6}$Ti$_{1/6}$ highlights a $J_C$ value of 1$\times \,10^{4} \text{A/cm}^2$ \cite{38}. In order to explore the enhancement of the $J_C$, the sample was annealed in flowing argon at 550 $\degree$ C for 6 hours. Post-annealing, the sample exhibits a reduced lattice parameter of a = b = c = 3.30(1) \text{\AA} (figure S1 in the SI). Compared to the as-cast sample, no noticeable changes are observed in the superconducting transition temperature or the upper critical field $H_{c2}(0)$ (Figures S2 and S3 in the SI \cite{25}). However, magnetization measurements as a function of magnetic field (MH, Figure S4 in the SI \cite{25}) show flux jumps in the annealed sample. The $J_C$ values were estimated from the surface points of the MH loops. Notably, the field dependence of $J_C$ demonstrates remarkably high values, exceeding the practical benchmark of $10^{5}$ $\text{A/cm}^2$ up to 5 T. We anticipate that systematic and optimized annealing protocols could further enhance the $J_C$ performance. A summary of the superconducting properties and a comparison with NbTi are provided in Table \ref{superconducting properties}. To gain deeper insights into the pinning mechanisms, we plotted the normalized pinning force density, $f_P = F_P/F_{Pmax}$ against the reduce field, h = $\text{H/H}_{irr}$ at different temperatures (H$_{irr}$ is the irreversible field) as shown in Figure \ref{fig4:MH}(c). $H_{irr}$ can be calculated empirically with the help of Kramer plot as shown in the inset of Figure \ref{fig4:MH}(c) J$_c^{0.5}$H$^{0.25}$ $\propto$ H$_{irr}-$H \cite{39, 40}. Experimentally, the pinning force density, $F_P$ can be calculated by multiplying the $J_C$ by the applied magnetic field. The scaling behavior of $f_P$ can be described by the Dew-Hughes model, $f_P \propto h^p(1-h)^q$, where the value of p and q describes the characteristics of the flux pinning \cite{41}. The data fits with the values p = 0.5 and q = 2, suggesting the dominance of surface pinning, implying that the grain boundaries act as effective pinning centres. The intrinsic severe lattice distortion is likely to introduce the pinning centre in the grain boundaries and therefore originate the high $J_C$ in the material.

\begin{table}[h!]
\caption{Superconducting state parameters of \sample. A comparison with NbTi (used in the International Thermonuclear Experimental Reactor) is shown here.}
\label{superconducting properties}
\begin{center}
\begin{tabular*}{1.0\columnwidth}{l@{\extracolsep{\fill}}llll}\hline\hline
Properties& unit& \sample& NbTi\\
\hline
\\[0.5ex]                                  
T$_{c}$& K& 8.0& 9.3 \cite{2}\\ 
H$_{c1}$(0)& mT& 68.32$\pm$ 0.2& 15 \cite{42}\\
H$_{c2}$(0)& T& 11.94 & 11.5 \cite{2}\\                      
$H_{c2}^{P}(0)$& T& 14.33& 17.2\\
$\xi _{GL}(0)$& \text{\AA}& 52.4$\pm$ 0.02& $\approx$50 \cite{43}\\
$\lambda _{GL}(0)$& \text{\AA}& 2850$\pm$ 10& NA\\
$\kappa _{GL}(0)$& &54.4 $\pm$ 2& NA\\
$\gamma$& mJmol$^{-1}$K$^{-2}$& 5.51 $\pm$ 0.05& 10.76 \cite{44}\\
$\beta$ & mJmol$^{-1}$K$^{-4}$& 0.22$\pm$ 0.0005& 0.15 \cite{45}\\
$\theta_{D}$& K& 206$\pm$ 1& 236 \cite{44}\\
$\lambda_{e-ph}$&  &0.86 $\pm$ 0.02& 0.82 \cite{46}\\
D$_{C}$(E$_{F}$)& states/ev f.u& 1.26 $\pm$ 0.1& 2.19 \cite{47}\\
$\omega_{ln}$&   &80.7 $\pm$ 0.1& 14.4 \cite{47}\\
$\Delta C_{el}/\gamma T_C$&   &2.30 $\pm$ 0.01& $\approx1.85$ \cite{44}\\
$J_C$& A/cm$^2$&  $0.14 \times10^6$& $\approx 10^6$\cite{1}
\\[0.5ex]
\hline\hline
\end{tabular*}
\par\medskip\footnotesize
\end{center}
\end{table}
The superconducting transition is a classic example of a second-order phase transition marked by a distinct jump in specific heat at the transition temperature. This discontinuity does not arise from changes in crystal structure, elastic properties or lattice parameters but rather from the formation of Cooper pairs, where the electrons near the Fermi level pair with opposite spins and momenta. We measured the temperature dependence of specific heat, C (T), down to 1.8 K and under applied magnetic fields ranging from 0 to 9 T, as shown in the inset of Figure \ref{Fig5:Spheat}(a). The specific heat data reveal a broad transition near T$_C$, which indicates some degree of disorder in the material. As the magnetic field increases, the jump in the specific heat shifts towards lower temperatures. In the normal state at low temperatures, the specific heat can be described by the relation:
\begin{equation}
    C (T) = \gamma T + \beta T^3
    \label{eqn1:Debye}
\end{equation}
Here, $\gamma T$ and $\beta T^3$ are the electronic and phononic contributions to the specific heat. By fitting the experimental data using eqn.\ref{eqn1:Debye} (see Figure \ref{Fig5:Spheat}(a)), we obtained the Sommerfeld coefficient $\gamma$ and lattice specific heat coefficient $\beta$. The fitted equation yields $\gamma$ = 5.51 (2)mJ/mol K$^2$ and $\beta$ = 0.2236 (4)mJ/mol K$^4$.  Using the value of $\beta$, the Debye temperature ($\theta _D$) was estimated as 206 K using the equation \cite{48}:
\begin{equation}
     \theta_D = \left(\frac{12\pi^{4}RN}{5\beta}\right)^{\frac{1}{3}}
\end{equation}
The value of $\theta _D$ agrees with the values obtained from BG fitting on the resistivity data. The electronic density of states at the Fermi level has been calculated as 1.26 states eV$^{-1}$ f.u.$^{-1}$  using the expression:
\begin{equation}
   D_c(E_F) = \frac{3 \gamma_n}{\pi^2 k_B^2 \left(1+ \lambda _{e-ph} \right)}
    \label{DOS}
\end{equation}
where \(k_\mathrm{B} = 1.38 \times 10^{-23} \, \text{J K}^{-1}\). Furthermore, the McMillan formula can be used to calculate the electron-phonon coupling constant according to the given expression:
\begin{equation}
    \lambda_{e-ph} = \frac{1.04 + \mu^*\ln(\theta_D/1.45 T_C)}{(1-0.62 \mu^*)\ln (\theta _D/1.45 T_C)}
    \label{lambda_e-ph}
\end{equation}
where, $\mu^*$ is the screened coulomb potential assumed as 0.13 for the intermetallics and the conventional superconductors. For the \sample \, the calculated $\lambda_{e-ph}$ = 0.86 is higher than the BCS limit, suggesting the possible strong coupling superconductivity in the material.

\begin{figure*}
    \centering
\adjustbox{width=2.0\columnwidth, center}
    {\includegraphics{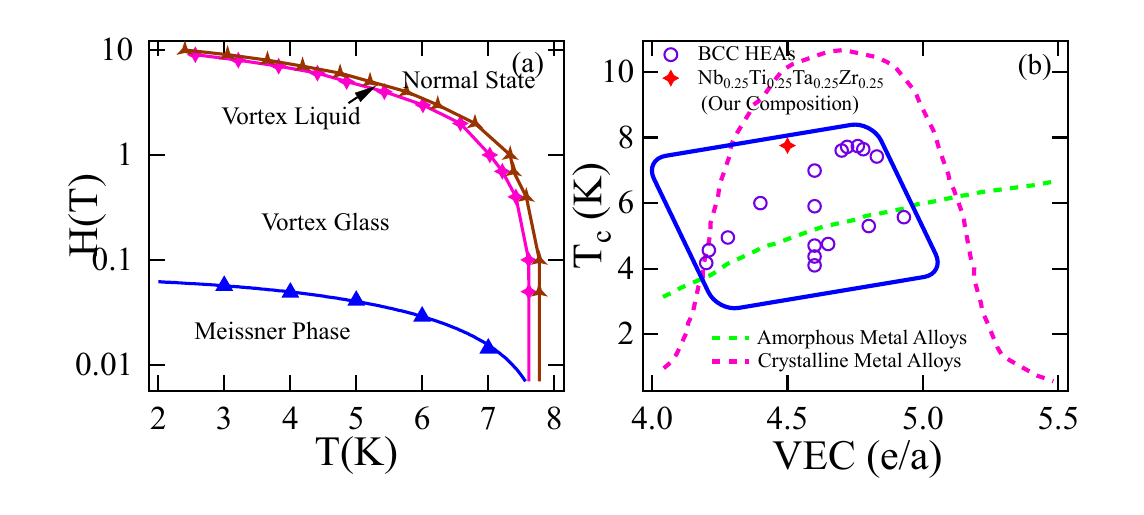}}
    \caption{(a) Vortex phase diagram of \sample \, showing transition from Meissner state to normal state. (b) Relationship between T$_C$ and VEC for various alloys \cite{27, 49, 50, 51, 52, 53, 54}.}
    \label{fig:vec}
\end{figure*}
We isolated the electronic specific heat (C$_{el}$) by subtracting the phononic contribution from the total specific heat to further investigate the pairing mechanism. Figure \ref{Fig5:Spheat}(b) shows the variation of normalized electronic-specific heat with temperature. We have analyzed the electronic-specific heat data using the $\alpha$ model \cite{55}. The $\alpha$ model is derived from the BCS theory but modified to include the multiple band's contribution, anisotropy, and strong coupling \cite{56}. We have observed a very high specific heat jump($\Delta C_{el}/\gamma T_C$) at the transition temperature $\approx$ 2.3, significantly higher than the BCS weak-coupling limit of 1.43. This large value of specific heat jump, together with the electron phonon coupling constant, categorizes the material as a strongly coupled superconductor. 
\begin{figure*}
   \centering
\adjustbox{width=2.0\columnwidth, center}
    {\includegraphics{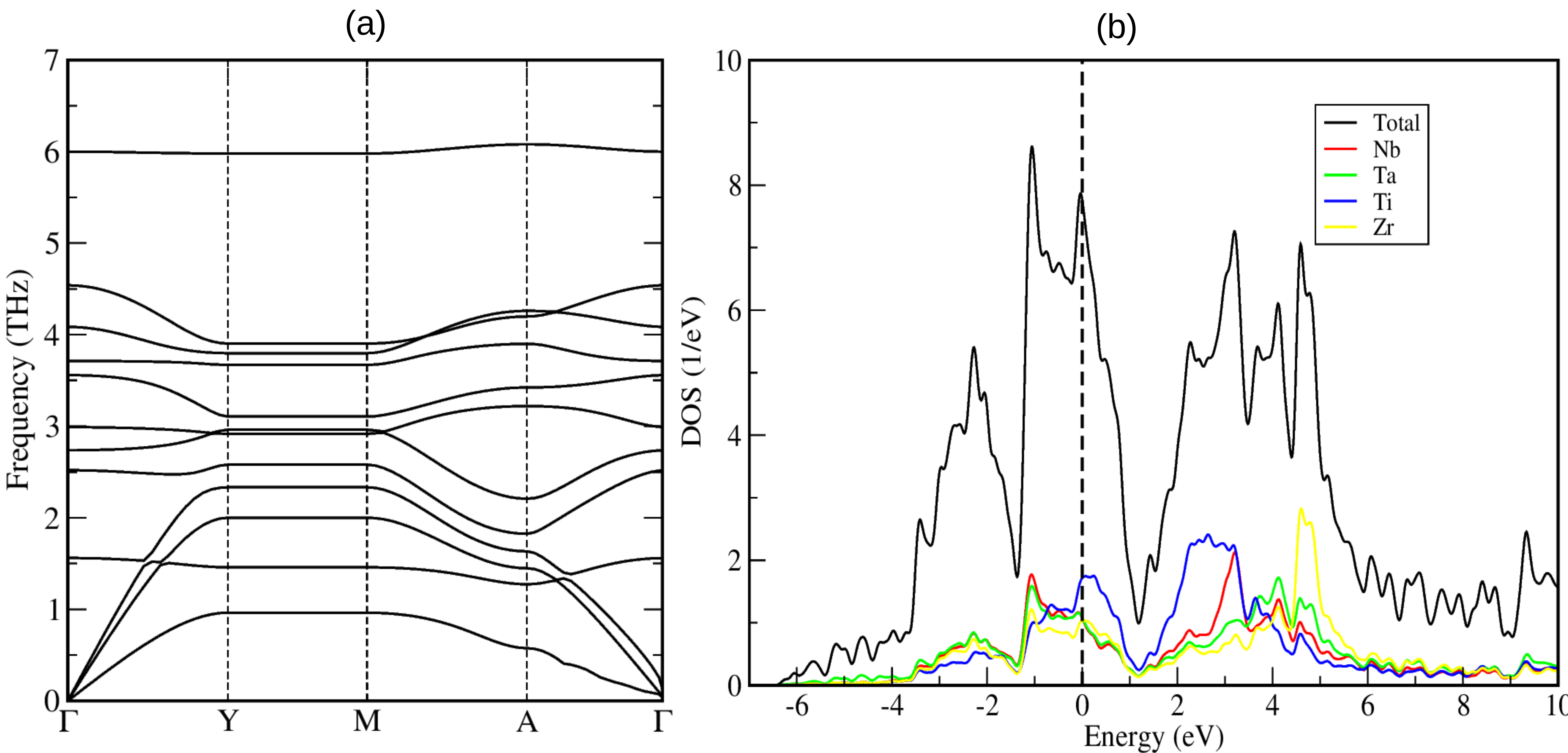}}
    \caption{(a) Calculated phonon dispersion of \sample \, along selected high-symmetry directions in the Brillouin zone, showing the absence of imaginary frequencies, thereby confirming the dynamical stability of the alloy. (b) Total density of states (DOS) and atom-resolved partial DOS projected onto the d-orbitals of each constituent transition-metal atom (Nb, Ta, Ti, and Zr), highlighting the dominant d-orbital contribution at the Fermi level that underpins the metallic nature and the potential role of multi-orbital interactions in the observed superconductivity.}
    \label{fig6: DOS}
\end{figure*}

The magnetic field dependence of the Sommerfeld coefficient, $\gamma$ (H), in the vortex state provides insight into the superconducting gap structure. In conventional isotropic s-wave superconductors, the normalized $\gamma$ (H), increases linearly with the applied field due to the contribution from quasiparticles located in vortex cores \cite{57}. In contrast, for superconductors with nodes in the gap, delocalized quasiparticles outside the vortex cores contribute significantly, leading to $\sqrt{H}$ dependence of $\gamma$ \cite{58, 59}. As shown in Figure \ref{Fig5:Spheat}(c), the electronic specific heat data is fitted with the equation:
\begin{equation}
    \frac{C_{el}}{T} = \gamma + \frac{A}{T}exp (-\frac{bT_C}{T})
\end{equation}
where the exponential term  $exp (-\frac{bT_C}{T})$ suggests nodeless superconductivity, consistent with the BCS theory. The obtained value of $\gamma$(H) is normalized with the normal state Sommerfeld coefficient and plotted against the normalized magnetic field represented in Figure \ref{Fig5:Spheat}(d). This clearly highlights that the normalized $\gamma$(H) deviated from the straight line behaviour and points to the existence of multiband superconductivity in the material. To analyze the dependency of $\gamma$ with magnetic field, the data is fitted with the equation: $\gamma \propto H^n$. The obtained value of n is $\approx$ to 0.55. This behavior is similar to the other well-studied multiband superconductors \cite{60, 61, 62, 63}.

In the strong coupling limit, where $(\frac{T_C}{\omega _{ln}} << 1)$, the normalized specific heat jump $\Delta C_{el}/\gamma T_C$ at the superconducting T$_C$, can be used to compute logarithmically averaged phonon frequency ($\omega _{ln}$), using the following equation:
\begin{equation}
    \frac{\Delta C_{el}}{ \gamma T_C}=1.43 \left[1+53\left(\frac{T_C}{\omega _{ln}}\right)^2
    \ln \left(\frac{\omega_{ln}}{3 T_C}\right)\right]
    \label{eq: strong-coupling}
\end{equation}
With the obtained value of $\omega _{ln}$ = 80.7 K, the value of superconducting gap, can also be obtained through the following equation
\begin{equation}
    \frac{2\Delta (0)}{k_B \ T_C} = 3.53 \left[1 + 12.5 \left( \frac{T_C}{\omega_{ln}} \right)^2 \ln \left( \frac{\omega_{ln}}{2 \ T_C} \right) \right]
    \label{BCSgap}
\end{equation}
From this, the ratio $\frac{\Delta (0)}{k_B \ T_C}$ is found to be 2.09, which closely matches the value obtained by the $\alpha$ model as illustrated in Figure \ref{Fig5:Spheat}(b). Additionally, the value of $\omega _{ln}$ is used in Allen-Dynes equation to estimate the electron-phonon coupling constant.
\begin{equation}
T_C = \frac{\omega_{\ln}}{1.2} \exp \left( -\frac{1.04(1+ \lambda_{e-ph})}{\lambda_{e-ph} - \mu^* (1+0.62 \ \lambda_{e-ph})} \right)
\label{AllenDynes}
\end{equation}
The resulting value, $\lambda _{e-ph}$ is 1.39, which indicates that the superconducting behavior of \sample \ is governed by the strong electron-phonon coupling.

The vortex phase diagram in Figure \ref{fig:vec}(a) is visualized with three distinct transition zones. The three zones are classified as the Meissner state, vortex glass state, and the vortex liquid state. At low magnetic fields and temperatures, the material exhibits perfect diamagnetism. As the field increases, vortices penetrate the superconductor and become immobilized due to pinning, forming a disordered but rigid structure known as vortex glass state. At even higher magnetic fields or temperatures, thermal fluctuations overcome pinning, allowing vortices to move freely, marking the vortex liquid state. Moreover, we have observed no other magnetization peak, which means elastic and plastic creep regions, often related to vortex motion and energy dissipation, are not present in our material. The irreversibility field, H$_{irr}$, defined at 10 $\%$ of the normal-state resistivity ($\rho _n$) separates the vortex glass state from the vortex liquid state \cite{64}. When the resistivity reaches 90 $\%$ of $\rho _n$, the system transitions to the normal state. Notably, the irreversibility line lies close to the upper critical field line, resulting in a narrow region of critical fluctuations. This feature is highly desirable for the practical applications of the superconducting material as it ensures robust superconductivity up to high magnetic fields \cite{65}. Turning to Figure \ref{fig:vec}(b), the plot demonstrates how the  T$_C$ of the crystalline solid solutions depends on the valence electron count (VEC) across different classes of materials, including crystalline transition metal alloys, amorphous alloys and BCC structured HEAs and MEAs \cite{27, 49, 50, 51, 52, 53, 54}. In BCC superconductors, T$_C$ tends to peak at specific VEC values, a phenomenon often observed in accordance with the empirical Matthias rule \cite{49}, which suggests optimal VEC ranges for high T$_C$. 

\begin{figure*}
   \centering
\adjustbox{width=2.0\columnwidth, center}
    {\includegraphics{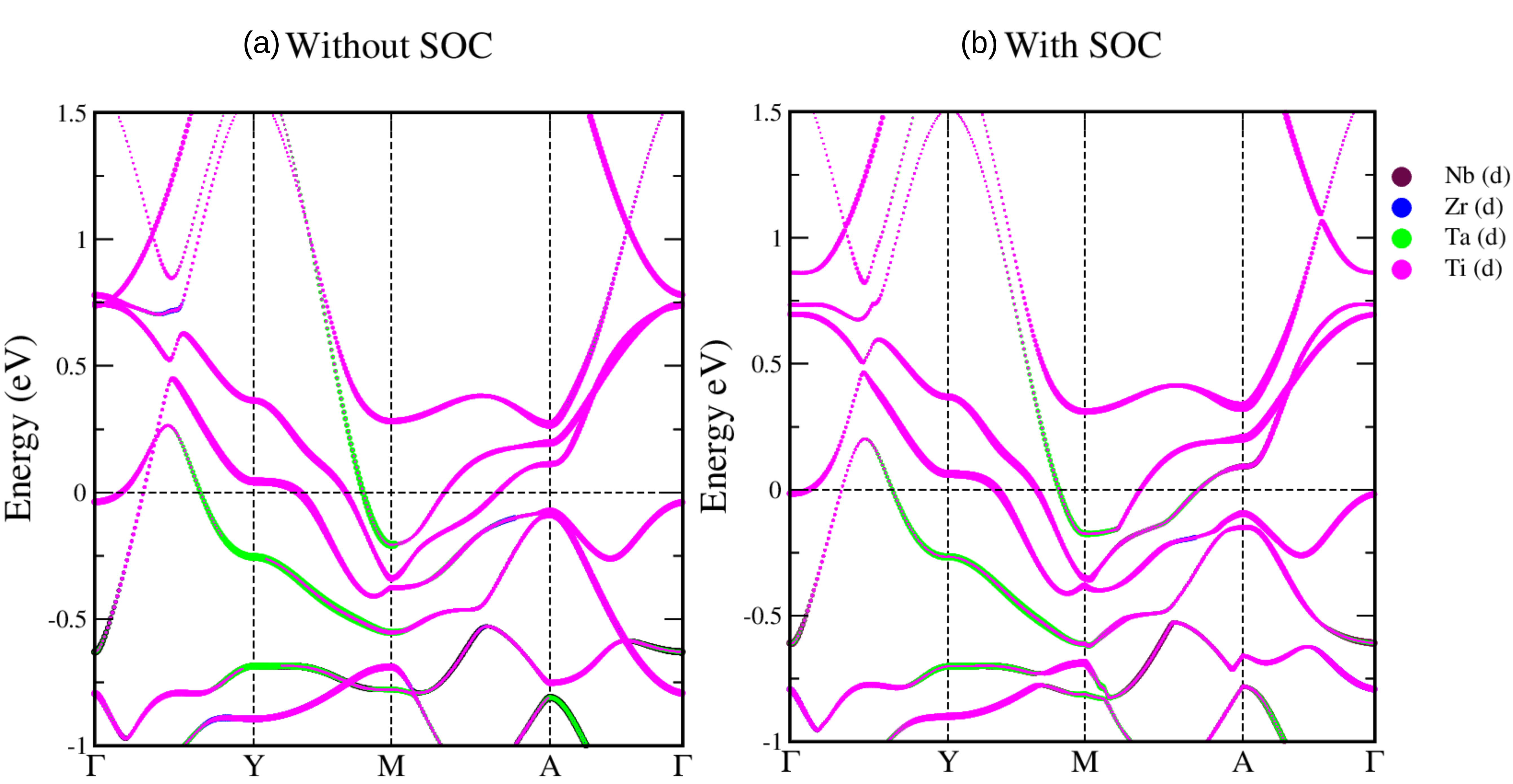}}
    \caption{Band dispersion of \sample \, along high symmetry directions calculated with and without spin-orbit coupling (SOC), with fatband representations highlighting the d-orbital contributions from each constituent atom. In the absence of SOC, several pronounced linear band crossings appear near the Fermi level, most notably between the $\Gamma$ and $Y$ points, and at the A point, between A and $\Gamma$ points, resembling Dirac-like dispersions. Upon inclusion of SOC, certain crossings evolve into well-defined gaps, while others (i.e.: between A and $\Gamma$) retain their degeneracy, indicating the presence of symmetry-protected states.}
    \label{fig7: Band}
\end{figure*}

The vibrational and electronic properties of the high-entropy alloy (HEA) $Nb_{0.25}Ti_{0.25}Ta_{0.25}Zr_{0.25}$, obtained via density functional theory (DFT) within the generalized gradient approximation (GGA), are presented in fig.~\ref{fig6: DOS}. These calculations offer fundamental insights into the structural stability, lattice dynamics, and the emergent electronic structure of the alloy, which are crucial for understanding its potential functional behavior. Figure~\ref{fig6: DOS}(a) shows the phonon dispersion spectrum along high-symmetry directions in the Brillouin zone ($\Gamma-Y-M-A-\Gamma$). The complete absence of imaginary phonon modes across the Brillouin zone confirms the dynamical stability of the relaxed structure, in line with expectations for a thermodynamically stable configuration. The phonon branches extend up to approximately 6 THz, with optical modes emerging above 3 THz. These optical branches exhibit relatively flat dispersions, indicating localized vibrational modes. These could be attributed to mass and force-constant disorder, both intrinsic features of HEAs. The suppression of phonon group velocities and the observed phonon localization are responsible for vibrational entropy and reduced thermal conductivity, driven by severe mass and bond-strength fluctuations among constituent elements. 

The electronic density of states (DOS), shown in Figure~\ref{fig6: DOS}(b), reveals a high spectral weight at the Fermi level ($E_F$), characteristic of a metallic system. This high DOS at $E_F$ indicates a significant electronic carrier density, favorable for electronic and thermoelectric applications. The partial DOS analysis identifies dominant contributions near $E_F$ from Ti and Zr d-orbitals, suggesting that these elements play a critical role in determining the low-energy electronic transport properties of the alloy. Figure 8 presents the calculated electronic band dispersion of \sample \, along high-symmetry directions, obtained with and without incorporating spin-orbit coupling (SOC), with fatband representations highlighting the d-orbital contributions from each constituent atom. 
The fatband analysis shows that states near the Fermi level are predominantly derived from Ti-d orbitals, with a smaller but noticeable contribution from Ta-d orbitals. In the absence of SOC, several pronounced linear band crossings emerge: (i) along the $\Gamma$ - $Y$ path at approximately $+0.2$ eV in the conduction band, (ii) at the A point around $-0.1$ eV in the valence band, and (iii) between the A and $\Gamma$ points at about $+0.35$ eV in the conduction band. These features resemble Dirac-like crossings that have been widely associated with symmetry-protected degeneracies in Dirac and Weyl semimetals \cite{66, 67, 68} Upon inclusion of SOC, the crossings along $\Gamma$ - $Y$ and at the A point are gapped, whereas the crossing between A and $\Gamma$ at $+0.35$ eV retains its degeneracy, indicative of a symmetry-protected band crossing \cite{69, 70, 71}. The persistence of this degeneracy despite SOC is often linked to rotational, inversion, or nonsymmorphic symmetries, as reported in symmetry-enforced Dirac semimetals \cite{72, 73, 74}. The appearance of such sharp, symmetry-resilient features in a chemically disordered medium-entropy alloy (MEA) is particularly noteworthy, since disorder generally broadens or eliminates such crossings \cite{75}. Further symmetry analysis, Berry curvature calculations, and experimental verification via angle-resolved photoemission spectroscopy (ARPES) will be essential to establish the precise topological character of these states. Nevertheless, the coexistence of Ti-d dominated Dirac-like crossings and strong intrinsic lattice distortion suggests that \sample \, may host robust, symmetry-protected metallic states or Dirac/Weyl-like quasiparticles in an entropy-stabilized, disordered environment, offering a promising route to realizing topology-driven transport phenomena in superconducting MEAs.

\section{Conclusions}
In summary, the medium-entropy alloy \sample \, having a BCC structure, shows superconductivity with a T$_C$ of 8 K and an upper critical field of $\approx$12 T. The critical current density exceeds the benchmark of 10$^5$ A/cm$^2$. This is attributed to effective surface pinning at the grain boundaries, likely due to the intrinsic severe lattice distortion. Specific heat analysis and the magnetic field dependence of the Sommerfeld coefficient suggest the presence of strong-coupling unconventional superconductivity in the material. First principle calculations confirm that the alloy is a dynamically stable metal, characterized by low-dispersion optical phonons and a high density of d$-$derived states at the Fermi level, dominated by Ti-d orbitals with secondary contributions from Ta-d. Notably, our DFT band-structure calculations reveal several Dirac-like band crossings near the Fermi energy, including symmetry-protected degeneracies that persist even in the presence of spin-orbit coupling. The survival of such sharp features in a chemically disordered matrix is remarkable, as disorder is generally expected to smear out or destroy these states, and hints at the  coexistence of topological electronic states with unconventional superconductivity 
in this system.
Our experimental and theoretical results suggest that the entropy-stabilized disordered alloys could be an ideal platform for exploring the interplay of severe lattice distortion and $J_C$, strong coupling unconventional superconductivity, and symmetry-protected topological phases, with potential implications in fusion reactors, quantum computing, and next-generation superconducting devices. 

\section{ACKNOWLEDGMENTS}
S. M. acknowledges the University of Bordeaux for the 'Professeur Invite' position through the Visiting Scholar Programme, supported within the framework of the France 2030 plan. S. M. also acknowledges the ANRF (previously SERB), Government of India, for the (SRG/2021/001993) Start-up Research Grant.

\section{Data Availability Statement}
The data that support the findings of this study are available from the corresponding author upon reasonable request.

\end{document}